\documentclass[aps,prl,preprint,superscriptaddress,showpacs]{revtex4}
\usepackage{graphics}
\usepackage{epsfig}
\usepackage{latexsym}
\usepackage{colordvi}
\usepackage{amsmath}
\usepackage{amssymb}

\def\lsim{\mathrel{\rlap{\lower3pt\hbox{\hskip1pt$\sim$}}
     \raise1pt\hbox{$<$}}} 
\def\gsim{\mathrel{\rlap{\lower3pt\hbox{\hskip1pt$\sim$}}
     \raise1pt\hbox{$>$}}}

\def\be{\begin{eqnarray}}\def\ba{\begin{eqnarray}}
\def\ee{\end{eqnarray}}\def\ea{\end{eqnarray}}
\def\ben{\begin{enumerate}}\def\bitem{\begin{itemize}}
\def\een{\end{enumerate}}\def\eitem{\end{itemize}}
\def\tr{{\rm tr}}

\begin{document}
\preprint{\parbox[b]{1in}{ \hbox{\tt IC/2008/038, INT-PUB-08/12} }}
\title{ Weak-Interacting Holographic QCD}

\author{Doron Gazit}
\email[E-mail: ]{doron.gazit@mail.huji.ac.il} 
 \affiliation{Institute for Nuclear Theory, University of Washington, Box 351550, 98195 Seattle,
 Washington, USA}

\author{Ho-Ung Yee}
\email[E-mail: ]{hyee@ictp.it} 
 \affiliation{ Abdus Salam International
Centre for Theoretical Physics, Strada Costiera 11, 34014, Trieste,
Italy} \vspace{0.1in}

\vspace{0.1in}

\date{\today}

\begin{abstract}

We propose a simple prescription for including low-energy
weak-interactions into the framework of holographic QCD, based on
the standard AdS/CFT dictionary of double-trace deformations. As our
proposal enables us to calculate various electro-weak observables
involving strongly coupled QCD, it opens a new perspective on
phenomenological applications of holographic QCD. We illustrate
efficiency and usefulness of our method by performing a few exemplar
calculations; neutron beta decay, charged pion weak decay, and
meson-nucleon parity non-conserving (PNC) couplings. The idea is
general enough to be implemented in both Sakai-Sugimoto as well as
Hard/Soft Wall models.

\end{abstract}

\maketitle
\newpage
\section{Introduction}
Holographic QCD is an attempt to construct an effective theory of
gauge invariant master fields in large N limit of QCD. One aspect of
it is the existence of an additional space-like dimension roughly
corresponding to energy scale of given observables. This is
plausible because the large N limit makes the theory of gauge
invariant operators classical, due to large N factorization, while
the concept of renormalization group survives in the limit.
Holographic QCD is a theory of gauge invariant fields in 5
dimensions reconciling with these aspects.

Experiences from the known correspondence between ${\cal N}=4$ SYM
in 4 dimensions and Type IIB string theory in $AdS_5\times S^5$ give
us an expectation that the would-be 5-dimensional holographic dual
theory of large N QCD becomes highly non-local in UV regime where
the corresponding large N QCD is asymptotically free, while we may
expect an approximate local theory in IR region as large N QCD
becomes strongly coupled. If our interest is only on low-energy
observables regarding strongly coupled QCD, it can be a worthwhile
approximation to model large N QCD by a local theory in 5
dimensions, as the current models of holographic QCD do. Presumably,
it is the UV asymptotic-free region where these local models fail to
describe large N QCD properly.

The existing holographic QCD models, like a top-down Sakai-Sugimoto
model \cite{Sakai:2004cn} or a bottom-up Hard/Soft Wall model
\cite{EKSS,PR,Karch:2006pv,ghoroku,BoschiFilho:2002ta},
 capture important aspects
of low energy QCD such as chiral symmetry and confinement, and
explain related experimental observables up to 20\%, which one might
expect from large N approximation. The previous calculations in the
framework concern only pure QCD sector and its electromagnetic
couplings (except Ref.\cite{Hambye:2005up} whose method is however
different from ours). As there are many important processes which
involve both weak-interaction and strongly-coupled low energy QCD,
it is pertinent to include effective weak-interactions in the
framework of holographic QCD, which will allow us a new tool for
estimating various hadronic weak processes.

The weak interactions are mediated by exchanges of heavy $W^{\pm}$
or $Z^0$ bosons. The tree level weak interactions are thus
proportional to the propagators of these bosons. In the low energy
limit, in which $q \ll M_{W^{\pm},Z^0}$, the interactions become
effective Fermi point-like vertices. We propose a simple
prescription for introducing these effective Fermi vertices of
weak-interaction in holographic QCD. In view of the original QCD,
this corresponds to perturbing the Lagrangian by the effective
four-fermi operators \be \Delta {\cal L}_{weak}={4 G_F \over
\sqrt{2}}\left(J_{W^+} J_{W^-}
+\cos^2\theta_W\left(J_{Z^0}\right)^2\right)= {4 G_F \over
\sqrt{2}}\left((J^1_L)^2+(J^2_L)^2+ \left(J^3_L-\sin^2\theta_W
J_Q\right)^2\right)\label{weakvertex} \ee where
$J_L^a=\sum_f\bar\psi^f_L{\sigma^a\over 2}\psi^f_L$ and $J_Q$ are
$SU(2)_L$ and electromagnetic currents respectively, and $\theta_W$
is the weak angle. When it is necessary, the currents in the above
may include leptonic part as well, which should be treated as an
external background.

In holographic QCD as a gauge/gravity correspondence, perturbing the
QCD Lagrangian would correspond to deforming boundary conditions of
5D fields near the UV boundary in a suitable way. Note that there is
a one-to-one map between single-trace\footnote{The "trace" here
means the trace over $SU(N)$ color indices.}
 operators in QCD and elementary
fields of holographic QCD in 5 dimensions. A 5D field $\phi(x,r)$,
where $r$ is the 5'th coordinate, behaves near the UV boundary
$r\to\infty$ as $ \phi(x,r)\sim c_1 r^{-\Delta_-}+c_2
r^{-\Delta_+}$ where $c_1$ is non-normalizable and $c_2$ is
a normalizable mode.
For the simplest case of perturbing with a single-trace
 operator $\cal O$,
\be \Delta {\cal L}=\int d^4x \, f(x){\cal O}(x)\quad, \ee it is
well-known that in the holographic dual theory the boundary
condition for the corresponding 5D field $\phi_{\cal O}(x,r)$  is
modified to be $c_1(x)=f(x)$, while the normalizable mode encodes
the expectation value of $\cal O$, that is, $c_2(x)=\langle {\cal
O}(x)\rangle$. However, the perturbation~(\ref{weakvertex}) that we
are interested in does not belong to this case; it consists of
products of two single-trace operators $J^\mu$'s. As the elementary
fields in holographic 5 dimensions only map to single-trace
operators, we can't simply introduce additional 5D fields for these
double-trace operators and deform their boundary conditions as
above.

There is an answer for this, first proposed for scalar operators
\cite{Witten:2001ua}, and generalized to global symmetry currents
\cite{Yee:2004ju,Marolf:2006nd}. Suppose we perturb by a general
polynomial of a single trace operator $\cal O$, \be \Delta {\cal
L}=\int d^4x\, F({\cal O})\quad. \ee The prescription is that this
modifies the boundary condition for the 5D field $\phi_{\cal O}$ to
be \be c_1(x)={\delta F({\cal O})\over \delta {\cal O}}\Bigg|_{{\cal
O}\to c_2(x)}\quad,\label{prescription} \ee where the normalization
should be such that $c_2$ is precisely equal to the expectation
value, while $c_1$ is the source for $\cal O$. We will be more
specific about precise normalization in our later discussions of
concrete examples.

Although the idea is general enough to be applicable to any model of
holographic QCD (see Ref.\cite{Erdmenger:2007cm} for a review), we
take two prototype examples, the Sakai-Sugimoto model  and the
Hard/Soft Wall model, to illustrate our proposal. In order to show
its usefulness, we analyze a few exemplar physical processes,
including the charged pion decay, neutron beta decay, and the
leading order parity non-conserving interaction between pions and
nucleons.
We stress that
the prescription we propose should be valid to
moderately higher energies before QCD asymptotic freedom sets in.

\section{Implementation}

{\it Sakai-Sugimoto model}

The Sakai-Sugimoto model is based on a string theory set-up of probe
$N_F$ $D8$ and ${\bar D8}$ branes in UV regime being joined at the
IR boundary of the background of color $N_c$ $D4$ branes,
geometrically realizing spontaneous chiral symmetry breaking
$U(N_F)_L\times U(N_F)_R\to U(N_F)_I$. The resulting configuration
can also be viewed as $N_F$ $D8$ branes with two asymptotic UV
regions, one for $U(N_F)_L$ and the other for $U(N_F)_R$. A
convenient choice of the 5'th coordinate is $Z$ ranging from one
boundary $Z=-\infty$ to the other $Z=+\infty$, where the IR regime
corresponds to near $Z=0$. Holographic dynamics of the chiral
symmetry resides in the world volume $U(N_F)$ gauge field $A(x,Z)$
on the $D8$ branes, whose values near two UV boundaries $Z\to\pm
\infty$ couple to the global $U(N_F)_L$ and $U(N_F)_R$ currents of
QCD respectively. We will focus on the two-flavor $N_F=2$ case in
the following, but its extension to $N_F>2$ is straightforward.

The part of the 5 dimensional action which is relevant for us is \be
S_{D8}=-\kappa\int d^4x\int dZ\,{\rm tr}\left[ {1\over
2}\left(1+Z^2\right)^{-{1\over
3}}F_{\mu\nu}^2+\left(1+Z^2\right)F_{\mu Z}^2\right]\quad,
\label{SSaction} \ee with $\kappa={\pi\over 4}f_\pi^2={\lambda
N_c\over 216\pi^3}\simeq 7.45\times 10^{-3}$ to best fit the
experiments for mesons, and the only scale in the model,
$M_{KK}\simeq 0.95\,{\rm GeV}$, is set to 1. Near the UV boundary
$Z\to \pm \infty$, the second term dominates, from which the
equation of motion asymptotes to $\partial_Z\left(Z^2\partial_Z
A_\mu\right)=0$, and we have \be A_\mu(Z) \sim A_\mu(\pm\infty) \pm
{A_\mu^{(1)}(\pm\infty)\over Z}+\cdots\quad,\label{SSexpansion} \ee
near $Z\to\pm\infty$. As usual, $A_\mu(+\infty)\equiv A_\mu^L$
couples to the $U(2)_L$ current $J^\mu_L$, and similarly for
$A_\mu(-\infty)\equiv A_\mu^R$ and $J^\mu_R$. The sub-leading
coefficients $A^{(1)}(\pm\infty)$ are proportional to the
expectation values of $J_{L,R}$, and to find the precise
normalization, we insert (\ref{SSexpansion}) in (\ref{SSaction}) and
take a variation with respect to sources $A^{L,R}$ to find the
expectation values of $J_{L,R}$. A quick calculation gives us \be
\eta_{\mu\nu}\langle J^\nu_{L,R}\rangle =- 2\kappa
A_\mu^{(1)}(\pm\infty)=\mp 2\kappa\, {\rm lim}_{Z\to\pm\infty}
\left(Z^2 F_{\mu Z}\right)\quad, \ee where the last expression is a
gauge invariant statement.

Having identified the normalized sub-leading expansion of the
holographic gauge field $A$, we are ready to implement the effective
weak interactions in the set-up. The perturbation (\ref{weakvertex})
contains both quark and lepton currents, and the quark currents are
nothing but of the chiral symmetry $U(2)_L\times U(2)_R$ as the
electro-weak interaction is a partial gauging of it. We treat the
lepton currents as external backgrounds in the present framework.
Recalling that $J_Q=J^3_L+J^3_R+{1\over 6}(J_L^{U(1)}+J_R^{U(1)})$
for quarks with $U(1)$ charge matrix being simply $2\times 2$
identity, the prescription (\ref{prescription}) for our double-trace
perturbation (\ref{weakvertex}) gives us the boundary conditions for
$A=A^a{\sigma^a\over 2}+A^{U(1)}{\bf 1}_{2\times 2}$
 near $Z\to\pm\infty$,
\be A^a_\mu(+\infty)&=&{8G_F\over \sqrt{2}}\left(-2\kappa\left(
Z^2 F_{\mu Z}^a\right)\Big|_{Z\to+\infty}+J^{a,{\rm lepton}}_{L\mu}\right)\quad,\label{amu}\\
A^a_\mu(-\infty)&=& 0\quad,\quad a=1,2\quad,\nonumber \ee from the
$W^\pm$ vertices, and from the $Z^0$ vertex we have \be {1\over 1-
\sin^2\theta_W}A^3_\mu(+\infty)
={-1\over\sin^2\theta_W}A^3_\mu(-\infty) ={-6\over\sin^2\theta_W}
A^{U(1)}_\mu(+\infty)={-6\over\sin^2\theta_W}A^{U(1)}_\mu(-\infty)=
K_\mu\quad,\nonumber \ee with \be K_\mu&=&{8G_F\over
\sqrt{2}}\left(-2\kappa Z^2 \left((1-\sin^2\theta_W)F_{\mu
Z}^3-{\sin^2\theta_W\over 6}
F^{U(1)}_{\mu Z}\right)\Bigg|_{Z\to+\infty}\right)\nonumber\\
&+& {8G_F\over \sqrt{2}}\left(-2\kappa \sin^2\theta_W
Z^2\left(F_{\mu Z}^3 +{1\over 6}F^{U(1)}_{\mu
Z}\right)\Bigg|_{Z\to-\infty}+{\rm
lepton\,\,currents}\right)\quad.\label{kmu} \ee

Normally, we are interested in effects which are first order in
$G_F$, and for that purpose the procedure of applying the above to
concrete examples can be simplified in the following way; first
solve the normalizable spectrum and wave-functions of the 4D particles of
interest without $G_F$, which has been done in the previous
literature. They
will contribute to the right-hand side of (\ref{amu}) and
(\ref{kmu}) and each mode then
entails a non-normalizable piece of first order in $G_F$, which
enters the calculation of various amplitudes of hadronic weak
processes.
Observe that the
non-normalizability issue of the $G_F$-perturbed wave-functions is
an order $(G_F)^2$ question.

{\it Hard/Soft Wall model}

We next implement our prescription in the Hard/Soft Wall model. The
Hard/Soft Wall model is a more phenomenological attempt by taking
simple AdS space-time with a hard/soft IR cutoff and introducing a
minimal set of fields relevant for the chiral symmetry and its
breaking. The action takes a form \be S_{5D}=\int d^4x\int
dz\,\sqrt{G}e^{-\Phi}\,{\rm tr} \left(-{1\over
4g_5^2}\left(F_{L,R}\right)^2 +|DX|^2+3|X|^2\right)\quad, \ee with
$U(2)_L\times U(2)_R$ gauge fields $A_{L,R}$, a bi-fundamental
scalar $X$, a dilaton $\Phi$, and $g_5^2={12\pi^2\over N_c}=4\pi^2$.
One virtue of the model is that a bare quark mass can be introduced
via the non-normalizable profile of $X$, while its normalizable mode
parameterizes chiral condensate. A popular coordinate system often
used is $ds^2={1\over z^2}\left(-dz^2+dx_\mu dx^\mu\right)$
 where the UV boundary is $z\to 0$. As the dilaton in the Soft
Wall model is $\Phi(z)=z^2\to 0$ near $z\to 0$, the UV dynamics of
the Hard and Soft Wall models is identical, and so is our
implementation of weak interactions via perturbed UV boundary
conditions.

Due to the expectation value $\langle X \rangle={1\over 2}(m_q
z+\sigma z^3){\bf 1}_{2\times 2} =v(z){\bf 1}_{2\times 2}$ which
breaks the axial part of the chiral symmetry $A\equiv{1\over
2}(A_L-A_R)$, there appear mixed kinetic terms between
$A_M=(A_\mu,A_z)$ and the phase part $P$ of $X=\langle X \rangle
e^{i P}$, which is typical in a Higgs mechanism, \be
S_{mixing}=-2\int d^4x dz\, {1\over z}\,{\rm tr}\left({1\over
g_5^2}\left(\partial_z A_\mu\right)\left(\partial^\mu A_z\right)+{2
v^2\over z^2}A_\mu\left(\partial^\mu P\right)\right) \quad, \ee
where the indices are contracted with flat metric
$\eta_{\mu\nu}={\rm diag}(+1,-1,-1,-1)$. This brings a slight
complication into the analysis of UV behavior of our 5D gauge
fields, but not in an essential way as we will see shortly. Note
that the isospin part of the chiral symmetry $V={1\over 2}(A_L+A_R)$
remains intact, and the asymptotic UV expansion from equation of
motion is easily found to be \be V_\mu \to V_\mu(0)+V_\mu^{(1)}
z^2+\cdots\quad, \ee and the usual interpretation of source and
expectation value goes as before.

To study the $(A_\mu,A_z,P)$ system clearly, it is convenient to
work in a 5D analog of $R_\xi$-gauge
 by adding the following gauge fixing term in the bulk \cite{Hong:2006ta,Hong:2007tf},
\be S_{g.f.}=-{1\over 2\xi}\int d^4x dz\,{1\over z} \,{\rm
tr}\left[\partial_\mu A^\mu -2\xi\left({1\over
g_5^2}z\partial_z\left({A_z\over z}\right) -{2v^2\over z^2}
P\right)\right]^2\quad, \ee which removes the mixing terms between
$A_\mu$ and $(A_z,P)$, up to an important left-over boundary term,
\be S_{bdry}={2\over g_5^2}\int d^4x\, {\tr}\left( {1\over z} A_\mu
\partial^\mu A_z \Big|_{z\to 0}\right)\quad.\label{bdry} \ee This
boundary term will be important later in discussing coupling of
external sources to the pions. The above gauge is especially useful
in identifying physical degrees of freedom in a transparent way.
Upon 4D reduction via KK mode expansion, one combination of $A_z$
and $P$ gives longitudinal components of massive vector mesons from
$A_\mu$, while the other contains the pions and the excited scalar
mesons. To separate longitudinal components of vector mesons from
the pions, we further take a unitary gauge $\xi\to\infty$, upon
which $(A_z,P)$ includes only physical pions and its excitations,
and allows us to replace $P$ with $A_z$ via \be {1\over
g_5^2}z\partial_z\left({A_z\over z}\right) ={2v^2\over z^2}
P\quad.\label{fp} \ee The resulting equation of motion for $A_z$ can
be solved for the pions and excited scalar mesons, with the UV
boundary condition $A_z\sim {\cal O}(z)$ for normalizability
\cite{Hong:2007tf}.

The action for $A_\mu$ in our gauge then becomes simply \be
S_{A}=\int d^4x dz\, {1\over z}\,{\rm tr} \left(-{1\over 2
g_5^2}(F_{\mu\nu})^2 +{1\over g_5^2}(\partial_z A_\mu)^2 +{4
v^2(z)\over z^2} A_\mu A^\mu\right)+S_{bdry}\quad,\label{Aaction}
\ee where indices are contracted with the flat metric
$ds^2=-dz^2+dx^2$, and its equation of motion along $z$ is \be
\partial_z\left({1\over z}\partial_z A_\mu \right)-
{1\over z}\partial_\nu F_{\nu\mu}-{4 g_5^2 v^2(z) \over z^3} A_\mu
=0\quad, \ee whose solution has a UV asymptotic \be A_\mu \sim c_1 z
K_1(g_5 m_q z)+c_2 zI_1(g_5 m_q z)\sim A_\mu^{(0)} +A_\mu^{(1)}z^2 +
{1\over 2}\left(\partial_\nu F^{(0)}_{\nu\mu}+g_5^2 m_q^2
A_\mu^{(0)}\right)z^2\log z+ \cdots\nonumber \ee which is
essentially same to the case without a Higgs breaking. We observe
that the UV behavior is not much affected by the symmetry breaking
through $\langle X\rangle\neq 0$, and we can apply the usual
dictionary of source and expectation value to $A_\mu^{(0)}$ and
$A_\mu^{(1)}$ respectively. It is also possible to prove this
assertion more rigorously by the standard procedure of holographic
renormalization \cite{Bianchi:2001kw}.
Observe that the previously
identified boundary term (\ref{bdry}) in our unitary gauge also
contributes to the expectation value as it becomes \be
S_{bdry}={2\over g_5^2}\int d^4x \, {\rm
tr}\left(A_\mu^{(0)}\left({1\over z}\partial^\mu
A_z\right)\Big|_{z\to 0}\right)\quad,\label{bdry2} \ee and there is
a similar term for $V={1\over 2}(A_L+A_R)$ too.

As we find that the usual gauge/gravity dictionary holds for the
$U(2)_L\times U(2)_R$ bulk gauge fields, it is straightforward to
implement our proposal for effective weak interactions. The precise
relation between $A^{(1)}_{L,R}$ and $\langle J_{L,R} \rangle$ from
the above discussion is \be \eta_{\mu\nu}\langle
J^\nu_{L,R}\rangle=-{2\over g_5^2}A^{(1)L,R}_\mu+{1\over
g_5^2}{1\over z}\partial_\mu A_z^{L,R}\Big|_{z\to 0} ={1\over
g_5^2}{1\over z}F^{L,R}_{\mu z}\Big|_{z\to 0}\quad, \ee where the
second contribution from (\ref{bdry2}) makes the final result gauge
invariant, which is important to include the cases with the pions
coming from $A_z\sim{\cal O}(z)$. Using this, the Fermi-interaction
via $W^\pm$ exchange corresponds to the boundary condition \be
A_{L\mu}^a(0) &=& {8G_F\over \sqrt{2}}\left({1\over g_5^2}{1\over z}
F^{L a}_{\mu z}\Big|_{z\to 0}
+J_{L\mu}^{a,leptons}\right)\quad,\quad a=1,2\quad,\nonumber\\
A_{R\mu}^a(0)&=& 0\quad,\label{hardbc} \ee and similarly for the
$Z^0$ exchange as in (\ref{kmu}).

The boundary conditions (\ref{amu}), (\ref{kmu}) and (\ref{hardbc})
are our main points in the paper. To find effects which are lowest
order in $G_F$, we first solve normalizable modes dynamics without
$G_F$, and via (\ref{hardbc}) non-normalizable perturbations of
linear order in $G_F$ are induced. We  point out that to leading
order in $G_F$, our prescription is
 equivalent to large N factorizing our
effective $JJ$ weak vertices and then calculating  chiral form
factors of hadrons.

\section{Exemplar Calculations}

{\it (1) Charged pion weak decay $\pi^+ \to \mu^+ \nu_\mu$}

For an illustrative purpose of showing how the prescription works,
we compute via our method a few simple hadronic weak processes
which involve only $W^\pm$ exchange, while the analysis of neutral
current phenomena and fitting the weak angle $\theta_W$ are left to
the future.

One of the cleanest observables from charged current exchange is the
decay of charged pions; more than 99\% of $\pi^+$ decay to $\mu^+
\nu_\mu$ via $W^+$-exchange, while the electron channel $\pi^+\to
e^+ \nu_e$ is helicity suppressed by ${m_e^2\over m_\mu^2}\sim
10^{-5}$. Computing the relevant pion coupling to the external
leptonic current $J_L^{+,\,lepton}=(J_L^{1}-iJ_L^2)^{lepton}=\bar
l_L \gamma^\mu \nu_L$ is in fact easy as the lepton current simply
acts as a source for the corresponding quark current through
(\ref{amu}) in the Sakai-Sugimoto model, or (\ref{hardbc}) for the
Hard/Soft Wall model. Expanding $A^a{\sigma^a\over 2}={1\over
\sqrt{2}}(A^+\sigma^-+A^-\sigma^+)$ for bulk fields \footnote{Note
that our convention is $A^+={1\over \sqrt{2}}(A^1-i A^2)$ with
$A=A^a{\sigma^a\over 2}$
 for bulk gauge fields,  while
$J^+=(J^1-i J^2)$ with $J^a=\bar\psi {\sigma^a\over 2}\psi$ for the
boundary currents. Therefore, the source-operator coupling becomes
$A^a J^a={1\over\sqrt{2}}(A^+ J^- +A^- J^+)$ in our convention.},
this is equivalent to having an external source \be
A^+(+\infty)&=&(4G_F)\, J_L^{+,\,lepton}=(4G_F)\, \bar l_L \gamma^\mu \nu_L \quad ({\rm Sakai-Sugimoto})\\
A_L^+(0)&=&(4G_F)\, \bar l_L \gamma^\mu \nu_L \quad ({\rm
Hard/Soft\,Wall})\quad. \ee

We consider the case of Sakai-Sugimoto model first. To identify how
pions interact with the above external source for chiral currents,
working in the $R_\xi$-gauge with unitary limit $\xi\to\infty$ again
seems most convenient.  By the similar procedure to reach
(\ref{Aaction}) in the Hard/Soft Wall models,
the final action in our unitary gauge becomes
\be S&=& -\kappa\int d^4x\int dZ \,{\rm tr}\left[{1\over
2}(1+Z^2)^{-{1\over 3}}
F_{\mu\nu}^2-(1+Z^2)(\partial_Z A_\mu)^2\right]\nonumber\\
 &&+ \int d^4x \,{\rm tr}\left(\partial_\mu \pi\right)^2
-f_\pi\, {\rm tr} \left[(\partial^\mu \pi )
A_\mu\Big|_{-\infty}^{+\infty}\right]\quad, \label{unitaryaction}
 \ee
where we used $A_Z
= {1\over \sqrt{\pi\kappa}}{1\over 1+Z^2} \,\pi(x)$ in our unitary gauge with
$\kappa={\pi\over 4}f_\pi^2$.
Note that the last term comes from the remaining boundary term after adding the gauge fixing
term as in (\ref{bdry}).
Mode
expansion of $A_\mu$ contains (pseudo) vector mesons as normalizable
modes plus  external sources for the chiral currents as
non-normalizable modes. These non-normalizable modes then  couple to
the pions only through the last boundary term that we are looking
for. The result in fact agrees with the old current algebra
expression. Expanding $\pi={1\over \sqrt{2}}(\pi^+\sigma^- +\pi^-
\sigma^+)$ and plugging in $A(+\infty)={1\over \sqrt{2}}A^+(+\infty)
\sigma^-={4G_F\over \sqrt{2}} (\bar l_L \gamma^\mu \nu_L)\sigma^-$,
we get the coupling \be \label{pion_decay_coupling} {\cal
L}_{\pi\bar l \nu}=-2 G_F f_\pi (\partial_\mu \pi^-)(\bar l_L
\gamma^\mu \nu_L)\quad. \ee

We expect the same result from the Hard/Soft Wall model as well.
Since in our gauge the pions reside in $A_z=\pi(x)f(z)$ with a
normalized profile $f(z)\sim z$, the previously identified boundary
term (\ref{bdry}) in the unitary gauge gives us the same type of
coupling between $\pi(x)$ and $A_L^+(0)=(4G_F)\, \bar l_L \gamma^\mu
\nu_L$.
To realize this, note that $A={1\over 2}(A_L-A_R)$, thus the
boundary term (\ref{bdry}) reproduces \be \left({1\over g_5^2
z}f(z)\Big|_{z\to 0}\right){\rm tr} \Big((\partial_\mu
\pi)(A_L^\mu-A_R^\mu)\Big|_{z\to 0} \Big) \equiv -f_\pi \,{\rm
tr}\Big((\partial_\mu \pi)(A_L^\mu-A_R^\mu)\Big|_{z\to 0}\Big)\quad,
\ee so that we have a near $z\to 0$ expansion of the pion profile
$A_z=\pi(x)f(z)$, \be f(z)=-(g_5^2 f_\pi ) z+\cdots
\quad,\label{fzexpand} \ee which will be useful for later
calculations involving the pions.
Recall that the coupling strength of the pions to  external current
sources, as appears in Eq.~(\ref{pion_decay_coupling}) is one of the
definitions of $f_\pi$. Using standard kinematics
\cite{Walecka_Book}, one uses the experimental decay rate to find
$f_\pi \approx 92 \, \mathrm{MeV}$. Thus, in weak-interacting
holographic QCD we successfully predict not only the operator
structure but also the value of the decay constant.

{\it(2) Neutron beta decay}

We next analyze a more complicated weak-process involving nucleons;
neutron beta decay to proton and  the first generation of leptons,
$n\to p^+  e^- \bar{\nu_e}$, with $\sim 100$\% branching ratio. As
we treat the lepton current $J_L^{-,\,lepton}=\bar{\nu_L}\gamma^\mu
e_L$ as external, the analysis is similar to the previous example,
that is, the lepton current will play a role of external source for
the chiral symmetry current \be
A^-(+\infty)&=&(4G_F)\, \bar{\nu_L} \gamma^\mu e_L \quad ({\rm Sakai-Sugimoto})\\
A_L^-(0)&=&(4G_F)\,  \bar{\nu_L} \gamma^\mu e_L \quad ({\rm
Hard/Soft\,Wall})\quad, \ee and we need how the nucleons couple to
the above external source of chiral currents. As a model
calculation, we proceed by adopting the recent effective field
approach to holographic baryons in
Ref.\cite{Hong:2007kx,Park:2008sp} for the Sakai-Sugimoto model, and
in Ref.\cite{Hong:2006ta} for the Hard Wall model. See also
Refs.\cite{Nawa:2006gv,Hata:2007mb,deTeramond:2005su,Pomarol:2007kr,Hata:2008xc,Hashimoto:2008zw,Kim}
for other approaches to holographic baryons.

One difference from the previous example of the pions is that the
resulting coupling would be momentum-dependent on general grounds.
In other words, the result should be phrased as a form factor
\cite{Polchinski:2001tt,Brod,Hong:2004sa,Grigoryan:2007wn}. The
resulting form factor in general can be rewritten as a sum over
infinite number of intermediate excitations, similar in spirit to
vector dominance \cite{Hong:2004sa}. The case of pions in the
previous example is special in this respect as they couple to the
bulk gauge fields only at the boundary through $S_{bdry}$
\footnote{See however Ref.\cite{Hov} for a 5D Chern-Simons term
induced 3-point anomalous coupling.}. As the neutron beta decay is a
3-body decay, the relevant momentum transfer is the invariant mass
square $q^2$ of the $e^-\bar{\nu_e}$ pair, which ranges $m_e^2 < q^2
< (m_n-m_p)^2$.
In practice however the relevant $q^2$ in our
case is too small compared to the QCD scale to make any notable
difference from simply using $q^2=0$ in numerical computations.

We work in the Sakai-Sugimoto model first.
In our previous unitary
gauge,
we solve the equation of motion for $A_\mu$ of a specific 4-momentum
$q^\mu$ with our prescribed boundary condition
$A_\mu(+\infty)={8G_F\over \sqrt{2}}J_{L\mu}^{lepton}$ and
$A_\mu(-\infty)=0$. Writing $A_\mu(x,Z)=A_\mu(Z) e^{-i q\cdot x}$,
the equation for $A_\mu(Z)$ in fact splits into two parts, one for
transverse and the other for longitudinal polarization,
and the final form of the solution is \be A_\mu =
A^T(q,Z)\left[\eta_{\mu\nu} -{q_\mu q_\nu \over
q^2}\right]A^\nu(+\infty) +{q_\mu\over q^2}(q\cdot
A(+\infty))\psi_+(Z)\quad,\label{finsol} \ee with $A^T(q,Z)$
satisfying
\be q^2 (1+Z^2)^{-{1\over 3}}
A^T+\partial_Z\left((1+Z^2) \partial_Z
A^T\right)=0\quad,\label{transeq} \ee
with the boundary condition
$A^T(q,+\infty)=1$, $A^T(q,-\infty)=0$, and $\psi_+(Z)={1\over 2}(1+{2\over \pi}
\tan^{-1}(Z))$. By computing how
holographic baryons couple to the above $q^2$-dependent bulk gauge
field, we read off nucleon interaction with an external source
$A_\mu(+\infty)$, which in our case is an external lepton current.

This is not the whole story however. There is an additional
important contribution coming from tree-level pion-exchange. This is
because pions couple to external currents by the last
term in (\ref{unitaryaction}), and
from
the interaction of holographic baryons to $A_Z$ in the bulk,
we can also obtain an axial coupling of pions to the nucleons, and
hence follows a coupling between external sources and the nucleons
via tree-level pion-exchange.

For a practical computation in the following, we should be specific
about a model of holographic baryons, for which we choose the model
in Ref.\cite{Hong:2007kx} as an example. The effective action of a
spin-$1\over 2$, $SU(2)$ doublet holographic baryon ${\cal B}$, as
the lowest spin state of  collective quantization of 5D
instanton-baryon on the $N_F=2$ $D8$ branes, is most easily written
in the conformal coordinate system $(x^\mu,w)$ with $
w=\int_0^{Z}\,{dZ' \over (1+Z'^2)^{2\over 3}}$, under which the 5D
baryon action doesn't involve spin connections, \be S_{baryon}=\int
d^4x\int dw \, \left[i\bar{\cal B} \gamma^M (\partial_M-i A_M){\cal
B} -m_B(w)\bar{\cal B}{\cal B} +(0.90) i\,\bar{\cal B}\gamma^{MN}
F_{MN} {\cal B}+\cdots\right]\,,\label{holbaction} \ee where the
position dependent mass term $m_B(w)$ can be obtained from the
energy of the $S^4$-wrapped $D4$ brane at the position $w$, whose
explicit form is not essential in our presentation, and the last
term is determined to reproduce the long-range tail of the
instanton-baryon solution \footnote{We performed the shift $N_c\to
N_c+2$ for the coefficient of the last term that was argued in
Ref.\cite{Hong:2007kx}.}. We then perform usual  KK reduction to 4D
by expanding ${\cal B}(x,w)=f_L(w)N_L(x)+f_R(w) N_R(x)$, where
$(N_L,N_R)\equiv N(x)$ is the 4D baryon mode  we identify as the
nucleons, and the profile functions $f_{L,R}(w)$ satisfy
a suitable eigen-value equation with the eigenvalue $m_N$ being the
mass of the nucleons. We have an important parity property
$f_R(-w)=f_L(w)$.
Once we find $f_{L,R}$,
we can insert our previous $A_\mu$ from (\ref{finsol}) and
$A_Z$  into the action (\ref{holbaction}) to find how
external sources and the pions couple to the nucleons.

After a straightforward calculation, the result is summarized by
\cite{Hong:2007kx,Hong:2007dq} \be {\cal L}_{N-external}=\bar
N\gamma^\mu {\cal V}_\mu N +\bar N\gamma^\mu\gamma^5 {\cal A}_\mu N
+\bar N \gamma^{\mu\nu}{\cal F}_{\mu\nu} N\quad,\label{Next1} \ee
with \be {\cal V}_\mu &=&{1\over 2}\int dw \,
\left(|f_L|^2+|f_R|^2\right) A_\mu(w)
+  (0.90) \int dw \,\left(|f_L|^2-|f_R|^2\right)\partial_w A_\mu(w)\quad,\nonumber\\
{\cal A}_\mu &=& {1\over 2}\int dw \, \left(|f_L|^2-|f_R|^2\right)
A_\mu(w)
+  (0.90) \int dw \,\left(|f_L|^2+|f_R|^2\right)\partial_w A_\mu(w)\quad,\nonumber\\
{\cal F}_{\mu\nu} &=& (0.54) i\, \int dw \, f_L f_R
\,F_{\mu\nu}(w)\quad,\label{Next2} \ee for the nucleon couplings to
external sources encoded in (\ref{finsol}), and \be {\cal L}_{N-\pi}
= -{g_A\over f_\pi} \,\bar N \gamma^\mu\gamma^5 (\partial_\mu \pi)
N\quad,\label{axial} \ee with \be g_A= 2\int dw\, |f_L|^2 \psi_0(w)
+4\cdot(0.90) \int dw\, |f_L|^2 \left(d\psi_0\over dw\right)\quad,
\label{ga} \ee for the pion-nucleon axial coupling. Numerical
evaluation gives $g_A\simeq 1.3$, which compares well with
experiment $g_A^{exp}=1.27$ \cite{Hong:2007kx}.

Although we can solve (\ref{transeq}) for $A^T$ numerically for
every $q^2$ of a problem to find $A_\mu(w)$ inserted in the above
result, it is a good approximation in our case to take the leading
small $q^2$ expansion as $q^2 \ll 1$,
which gives a simple formula
\be {\cal V}_\mu={1\over
2}A_\mu(+\infty)\quad,\quad {\cal A}_\mu={g_A\over 2}
A_\mu(+\infty)\quad, \ee with the exactly same $g_A$ as in
(\ref{ga}), and numerically we have \be {\cal F}_{\mu\nu}={1\over
2}\cdot(0.90)i \left(\int dw\,f_L f_R(w)\right) F_{\mu\nu}(+\infty)
\simeq (0.42) i  \,F_{\mu\nu}(+\infty)\quad. \ee Interestingly, it
can be easily  checked that the contribution from the tree-level
pion exchange has the similar
shape to the one from the ${\cal A}_\mu$ above, \be {\cal
L}_{\pi-exchange}=-{g_A\over 2}\,{q_\mu q_\nu\over q^2}\, \bar N
\gamma^\mu\gamma^5 A^\nu(+\infty) N\quad,\label{piexc} \ee so that
the sum of the all contributions is \be {\cal
L}_{N-external}&=&{1\over 2}\bar N\gamma^\mu A_\mu(+\infty) N
+{g_A\over 2}\,\left(\eta_{\mu\nu}-{q_\mu q_\nu\over q^2}\right)\,
\bar N \gamma^\mu\gamma^5 A^\nu(+\infty) N\nonumber\\
&+&(0.42) i\,\bar N \gamma^{\mu\nu}F_{\mu\nu}(+\infty) N +{\cal
O}(q^2)\quad. \ee Expanding $A_\mu(+\infty)={1\over\sqrt{2}}
A^-(+\infty)\sigma^+= {4G_F\over\sqrt{2}} \,( \bar{\nu_L} \gamma^\mu
e_L)\sigma^+$ and $N=(p,n)^T$, we finally have \be
\label{sagai_sugimoto_int} {\cal L}_{\bar n p
e^-\bar\nu_e}={4G_F\over\sqrt{2}}\left[{1\over 2}\, (\bar
n\gamma_\mu p)+{g_A\over 2}\,\left(\eta_{\mu\nu}-{q_\mu q_\nu\over
q^2}\right)\, (\bar n\gamma^\nu \gamma^5 p) -(0.84)\,q^\nu \,(\bar n
\gamma_{\mu\nu}p) \right] ( \bar{\nu_L} \gamma^\mu e_L)\,, \ee for
the effective vertex of the neutron beta decay. Note that the
Sakai-Sugimoto model does not have a bare quark mass by construction
and the pions are massless \footnote{See
Ref.\cite{Bergman:2007pm,Aharony:2008an} for attempts to introduce
quark mass in the Sakai-Sugimoto model.}. In any realistic
application of the above result,  we therefore should replace the
$q^2$-pole in the above with $m_\pi^2$. This unsatisfactory feature
will be absent in the Hard Wall model however, as we can introduce a
bare quark mass in the set-up.

We next turn to the analysis in the Hard Wall model. As the
procedure is identical to the previous one in the Sakai-Sugimoto
model, we will be brief in explaining the intermediate steps while
showing the results.
For our
purpose we will work in the holographic baryon model of
Ref.\cite{Hong:2006ta} for the Hard Wall model as it is most suitable to see
how holographic baryons couple to external chiral currents.
The holographic baryon action for the Hard Wall model reads as \be
S_N&=&\int d^4x \int dz \sqrt{G_5}\left[ i\bar N_1 \Gamma^M D_M N_1
+ i\bar N_2 \Gamma^M D_M N_2
-{m_5}\bar N_1 N_1+{m_5}\bar N_2 N_2\right]\nonumber\\
&+& \int d^4x \int dz \sqrt{G_5}\left[-g\bar N_1 X N_2 - g \bar N_2 X^\dagger N_1\right]\nonumber\\
&+& \int d^4x \int dz \sqrt{G_5}\left[(iD)(\bar
N_1\Gamma^{MN}F_{MN}^L N_1 - \bar N_2\Gamma^{MN}F_{MN}^R
N_2)\right]\quad,\label{Naction} \ee where $N_1$ ($N_2$) is a
doublet under $SU(2)_L$ ($SU(2)_R$).
The 5D Dirac mass $m_5$ is related to the scaling dimension $\Delta$
of the QCD 3-quark nucleon operator by $m_5=\left(\Delta-2\right)$,
and for simplicity we take $\Delta={9\over 2}$.
The second line gives us the observed nucleon mass $m_N=0.94$ GeV
when the chiral symmetry is broken $\langle X\rangle \neq 0$.
For
the popular values $z_m=(330 \,{\rm MeV})^{-1}$, $\sigma=(311\, {\rm
MeV})^3$ and $m_q=2.34\,{\rm MeV}$, this determines the coupling $g$
to be $g=9.18$. We leave the coefficient $D$ as a free parameter.
For these parameters, we have
$f_\pi=0.084 \,{\rm GeV}$ and the pion-nucleon axial coupling
is numerically $g_A=(0.33+1.02 D)$ where
$D$ is in units of $({\rm GeV})^{-1}$.


As the tree-level pion-exchange contribution is given by
(\ref{piexc}) except the $q^2$-pole replaced by $m_\pi^2$, it is an
order ${\cal O} (q^2)$ effect which is negligible.
By essentially same procedures as in the Sakai-Sugimoto case, our final
effective interaction for the neutron beta decay in the Hard Wall model is found to be \be
\label{hard_soft_int} {\cal L}_{\bar n p
e^-\bar\nu_e}={4G_F\over\sqrt{2}}\left[{1\over 2}\, (\bar
n\gamma_\mu p)+{1\over 2}\left(0.33+1.02 D \right)\, (\bar
n\gamma_\mu \gamma^5 p) -(0.48 D)\,q^\nu \,(\bar n \gamma_{\mu\nu}p)
\right] ( \bar{\nu_L} \gamma^\mu e_L). \ee
The neutron beta decay rate depends linearly on the axial form
factor at zero energy, i.e. $g_A$. In fact, $g_A$ is calibrated
using this process. Since the Sagai-Sugimoto model overpredicts
$g_A$ by about $2\%$, it would also underpredict the neutron
half-life by the same ratio.

In order to estimate the prediction of the Hard/Soft Wall model, we
need to fix $D$. In principle, this could be done by using the
physical value of either the axial form factor or the weak magnetism
form factor. In the Sagai-Sugimoto model, these two form factors are
successfully predicted as can be seen in
Eq.~(\ref{sagai_sugimoto_int}). In the case of Hard Wall model, we
find a tension between the two, that is if we choose to fit $D\simeq
0.94$ by $g_A^{exp}=1.27$, then the magnetic form factor is too
small by 50\%. There is however a chance for improving the bottom-up
Hard Wall model, for which we hope to come back in the future
\cite{doron}.

{\it (3) Parity non-conserving (PNC) pion-nucleon coupling}

Our final example is something whose theoretical estimate has been
difficult to obtain with other conventional tools. It is also a
first non-trivial example which does not involve external leptons,
so that the full aspect of our proposal should be used for its
calculation.
We are interested in the parity-violating couplings of mesons,
especially the pions, to the nucleons induced by weak-interactions.
For an illustrative purpose, we focus here only on the charged
pion-nucleon coupling from a $W^\pm$-exchange, and a more complete
study of the problem will be presented elsewhere \cite{doron}.

As before, the procedure of applying (\ref{amu}) and (\ref{hardbc}) to first
order in $G_F$ starts by first solving, without $G_F$, the relevant
normalizable modes for the particles involved in the process one is
trying to study.
Some
of these normalizable modes, such as mesons, come from the bulk
gauge fields of chiral symmetry, and the right-hand side of
(\ref{amu}) or (\ref{hardbc}) is non-vanishing for them in general.
These
modes are then accompanied by the induced boundary values in the
left-hand side of (\ref{amu}) or (\ref{hardbc}), which would act
just like external sources for the chiral currents. From studying
how external sources couple to the other particles of interest, we
can obtain the $G_F$-induced interactions between the original modes
and the other particles.

It is the pion in our case which carries an induced external source
by (\ref{amu}) or (\ref{hardbc}), and how this external source would
couple to the nucleons is described in the previous sections. In the
Sakai-Sugimoto model, the pions are sitting in $A_Z$ as before
and from (\ref{amu}), this entails a source \be A_\mu(+\infty)=
-{8G_F\over\sqrt{2}}\sqrt{4\kappa\over\pi}\left(\partial_\mu
\pi\right) =-{8G_F\over\sqrt{2}}
f_\pi\left(\partial_\mu\pi\right)\quad. \ee
We have the same result
in the Hard/Soft Wall model.
Once we find the induced
source, we can come straight to our previous results of nucleon
couplings to external sources; (\ref{Next1}) and (\ref{Next2}) for
the Sakai-Sugimoto model and similar expressions for the
Hard Wall model. As we may need a general momentum transfer $q^\mu$
between the nucleons and the pions, we should solve the necessary
$A^T(q^2)$ profile function numerically for a given $q^2$.
For the $q^2\to 0$ limit however, we have a simplification as before
to get \be {\cal L}^{weak}_{ N-\pi}=-2 G_F f_\pi \left( (\bar
p\gamma^\mu n) \left(\partial_\mu\pi^+\right)
+g_A\left(\eta_{\mu\nu}-{q_\mu q_\nu\over q^2}\right)(\bar p
\gamma^\mu\gamma^5 n) \left(\partial^\nu\pi^+\right)
\right)\,+\,{\rm h.c.}\quad, \ee for the Sakai-Sugimoto model, and
similarly for the Hard Wall model except the pion-exchange term.
Note that
${\cal F}_{\mu\nu}=0$ in our case as the source is purely
longitudinal. Only the first piece in the above corresponds to the
parity non-conserving (PNC) coupling, while the second term is a
small correction to the usual $g_A$. The first term is identical to
the contribution of pion exchange to the PNC coupling, as achieved
by current algebra and chiral perturbation theory \cite{PNC}.
Although the above pion-nucleon coupling in $q^2\to 0$ limit is
largely dictated by chiral algebra alone, its non-trivial
$q^2$-dependence when we consider finite $q^2$ is beyond the ability
of the chiral algebra. Our framework can give predictions for that.
Our method can also provide predictions for the coupling of other
excited mesons such as $\rho$, $\omega$, {\it etc}, whose details
will be reported soon elsewhere \cite{doron}.

\section{Discussion}

We have outlined a proposal for including effective weak
interactions in the framework of holographic QCD. By construction,
this prescription should be applicable up to energies of a few GeV.
In order to demonstrate the implementation of the prescription, we
chose two specific models, the Sagai-Sugimoto and Hard/Soft Wall
models. The use of our method in three low- energy processes, namely
charged pion decay, neutron beta decay, and PNC nuclear-meson
couplings, showed the usefulness and strength of our prescription.

Contrary to other theoretical tools to calculate such reactions,
e.g. chiral perturbation theory, the current approach not only
recovers the operator structure of previous methods, but also gives
a quantitative estimate, up to about 10-20\%, of the coupling
constants. In addition, our weak-interacting holographic QCD can be
extended to energies higher than the QCD scale, which is not
assessable in other methods.

Of course, there are many things to be done to improve the model. We
hope that our first step will stimulate further study and possible
refinement of our present proposal.

\subsection*{Acknowledgments}

We thank Institute for Nuclear Theory (INT) at the University of
Washington for the program "From Strings to Things: String Theory
Methods in QCD and Hadron Physics", where this work was initiated,
and H.U.Y. thanks the organizers for hospitality.
 D. G. is supported by DOE grant number DE-FG0200ER41132.

\end{document}